\documentclass[twocolumn,showpacs,preprintnumbers,prb]{revtex4}

\usepackage{graphicx}
\usepackage{dcolumn}
\usepackage{bm}
\usepackage{psfig}

\begin{document}

\title{Method for direct observation of coherent quantum oscillations in a
superconducting phase qubit. Computer simulations.}
\author{Ya. S. Greenberg}

\address{Novosibirsk State Technical University, 20 K. Marx Ave.,
630092 Novosibirsk, Russia}

\date{\today}

\begin{abstract}Time-domain observations of coherent oscillations between
quantum states in mesoscopic superconducting systems have so far
been restricted to restoring the time-dependent probability
distribution from the readout statistics. We propose a method for
\emph{direct }observation of Rabi oscillations in a phase qubit.
The external source, typically in GHz range, induces transitions
between the qubit levels. The resulting Rabi oscillations of
supercurrent in the qubit loop are detected by a high quality
resonant tank circuit, inductively coupled to the phase qubit.
Here we present the results of detailed computer simulations of
the interaction of a classical object (resonant tank circuit) with
a quantum object (phase qubit). We explicitly account for the back
action of a tank circuit and for the unpredictable nature of
outcome of a single measurement. According to the results of our
simulations the Rabi oscillations in MHz range can be detected
using conventional NMR pulse Fourier technique.
\end{abstract}

\maketitle

\section{Introduction}

As is known the persistent current qubit (phase qubit) is one of
the candidates as a key element of a scalable solid state quantum
processor.\cite{Mooij1,Orlando} The basic dynamic manifestations
of a quantum nature of the qubit are macroscopic quantum coherent
(MQC) oscillations (Rabi oscillations) between its two basis
states, which are differed by the direction of macroscopic current
in the qubit loop.

Up to now Rabi oscillations in the time domain
\cite{Nakamura,Esteve} or as a function of the perturbation power
\cite{Han} have been detected indirectly through the statistics of
switching events (e.g., escapes into continuum). In either case
the probability $P(t)$ or $P(E)$ was obtained and analyzed to
detect the oscillations.

 More attractive  is a
direct detection of MQC oscillations through a weak continuous
measurement of a classical variable, which would implicitly
incorporate the statistics of quantum switching events, not
destroying the quantum coherence of the qubit at the same
time.\cite{Averin,Korotkov,Korotkov1}

 In this paper we describe an approach which allows a direct detection of
 MQC oscillations of macroscopic current flowing in a loop of a phase qubit.
 To be specific, we
will use the example of three-Josephson-junction (3JJ)
small-inductance phase qubit (persistent current qubit\cite
{Orlando}) where level anticrossing was already
observed.\cite{Mooij}

\begin{figure}[tbp]
\includegraphics[width=7cm]{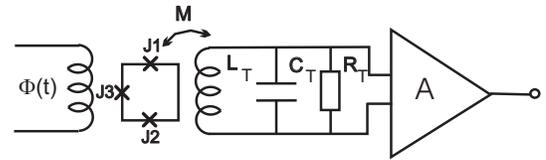}
\caption{Phase qubit coupled to a tank circuit.} \label{fig1}
\end{figure}

In our method a resonant tank circuit with known inductance
$L_{T}$, capacitance $C_{T}$, and quality $Q_T$ is coupled with a
target Josephson circuit through the mutual inductance $M$
(Fig.~\ref{fig1}).

 The method was successfully applied to a
3JJ qubit in classical regime\cite {APL7} when the hysteretic
dependence of qubit energy on the external magnetic flux was
restored in accordance to the predictions of
Ref.~\onlinecite{Orlando}.

Here we extend the approach which is described in our earlier
paper.\cite{Greenberg1} We explicitly account for the back action
of a tank circuit and for the unpredictable quantum mechanical
nature of outcome of a single measurement. According to the
results of our simulations the Rabi oscillations in MHz range can
be detected using conventional NMR pulse Fourier technique.


\section{Quantum dynamics of 3JJ flux qubit}
\label{3dd-dyn}

Quantum dynamics of this qubit has been studied in detail in
Ref.~\onlinecite{Orlando}. The qubit consists of a loop with three
Josephson junctions.  The loop has very small inductance,
typically in the pH range. It insures effective decoupling of
qubit from external environment. Two Josephson junctions have
equal critical current $I_\mathrm{c}$ and capacitance $C$, while
the critical current and capacitance of a third junction is a
little bit smaller, $\alpha I_\mathrm{c}$, $\alpha C$, where
$0.5<\alpha<1$. If the Josephson coupling energy
$E_\mathrm{J}=I_\mathrm{c}\Phi_0/2\pi$, where $\Phi_0=h/2e$ is a
flux quantum, is much more than the Coulomb energy $E_C=e^2\!/2C$,
then the phase of a Cooper pair wave function is well defined. As
was shown in Refs.~\onlinecite{Orlando, Mooij1} in the vicinity of
$\Phi=\Phi_0/2$ this system has two quantum stable states with
persistent circulating current of opposite sign. Therefore, the
persistent current qubit can be described by following two-level
Hamiltonian:
\begin{equation}\label{H_q}
H_q=-\hbar(\varepsilon(t)\sigma_z+\Delta\sigma_x),
\end{equation}
where $\sigma_z$ and $\sigma_x$ are Pauli spin operators.

Hamiltonian $H_q$ is written in the flux basis: the basis of
localized (left, right) states, so that two eigenvectors of
$\sigma_z$ correspond to the two classical states with a left or a
right circulating current. $\Delta$ is the tunnel frequency and
$\varepsilon(t)$ is in general a time dependent bias which is
controlled by externally applied flux
\begin{equation}\label{flux}
\Phi(t) =\Phi _{x}+\Phi _{ac}(t),
\end{equation}
where $\Phi _{x}$ is a time independent external flux, $\Phi
_{ac}(t)$ is a monochromatic high frequency signal from the
external source. According to (\ref{flux}) we write
$\varepsilon(t)$ in the form:
\begin{equation}\label{eps}
\varepsilon(t) =\varepsilon_0+\widetilde{\varepsilon}(t).
\end{equation}

In the absence of time dependent flux
($\widetilde{\varepsilon}(t)=0)$ the eigenstates of Hamiltonian
$H_q$ are
\begin{equation}\label{energ}
E_\pm=\pm\hbar\Omega\equiv\pm\hbar\sqrt{\varepsilon_0^2+\Delta^2},
\end{equation}
where $\varepsilon_0=E_Jf\lambda(\alpha,g)/\hbar$,
$f=\Phi_x/\Phi_0-\frac{1}{2}\ll1$. The explicit dependence of
$\lambda(\alpha,g)$ on qubit parameters $\alpha$ and
$g=E_\mathrm{J}/E_C$ has been found in
Ref.~\onlinecite{Greenberg}. The flux states are the eigenstates
of a macroscopic current which circulates in a qubit loop. Within
a harmonic approximation \cite{Greenberg} it is not difficult to
find current operator in the flux basis: $\hat{I_q
}=I_C(\lambda(\alpha,g)/2\pi)\sigma_z$.

Hamiltonian $H_q$ can be written in eigenstate basis as
$H_q=-\hbar\Omega\sigma_Z$ with eigenfunctions $\Psi_\pm$:
$H_0\Psi_\pm=E_\pm\Psi_\pm$. The stationary state wave functions
$\Psi_\pm$ can be written as the superpositions of the wave
functions in the flux basis $\Psi_L, \Psi_R$, where $L$, $R$ stand
for the left and right well respectively:
$\Psi_\pm=\emph{a}_\pm\Psi_L+\emph{b}_\pm\Psi_R$,
\begin{equation}  \label{a,b}
a_\pm=\frac{\Delta}{\sqrt{2\Omega(\Omega\mp\varepsilon_0)}},b_\pm=\frac{%
\varepsilon_0\mp\Omega}{\sqrt{2\Omega(\Omega\mp\varepsilon_0)}}.
\end{equation}
 The transformation from
flux basis to eigenstate basis is performed with the aid of
rotation matrix $R(\xi)=\exp(i\sigma_Y\xi/2)$, where
$\cos\xi=\varepsilon_0/\Omega$, $\sin\xi=\Delta/\Omega$.
Therefore, the current operator in eigenstate basis is

\begin{equation}  \label{curr}
\widehat{I_q}=I_C\frac{\lambda(\alpha,g)}{2\pi\Omega}\left(\varepsilon_0\sigma_Z
-\Delta\sigma_X\right).
\end{equation}

 The high
frequency excitation applied to the qubit induces the transitions
between two levels which result
in a superposition state for the wave function of the system: $%
\Psi(t)=C_+(t)\Psi_++C_-(t)\Psi_-$. The coefficients $C_\pm(t)$
can be obtained from the solution of time dependent
Schr\"{o}dinger equation with proper initial conditions for
$C_\pm(t)$. From (\ref{curr}) we obtain the average current in the
superposition state $\Psi(t)$:

\begin{eqnarray}\label{avcurr}
\langle\Psi(t)|\widehat{I}_q|\Psi(t)\rangle&=&I_C\frac{\lambda(\alpha,g)}{2\pi}\frac{\varepsilon_0}{\Omega}
\left(|C_-(t)|^2-|C_+(t)|^2\right) \nonumber\\
&&-I_C\frac{\lambda(\alpha,g)}{\pi}\frac{\Delta}{\Omega}Re(C_+(t)C_-^*(t)).
\end{eqnarray}
 Notice that at the
degeneracy point ($f=0$) the low frequency part of the average
current, which is given by first term in (\ref{avcurr}), vanishes.

 Below we define the density matrix elements
$\rho_{00}=|C_-(t)|^2$, $\rho_{11}=|C_+(t)|^2$,
$\rho_{10}=C_+(t)C^\ast_-(t)$, $\rho_{01}=\rho^\ast_{10}$,
$\rho_{00}+\rho_{11}=1$ which are related to spin operators:
$\langle\Psi(t)|\sigma_Z|\Psi(t)\rangle=2\rho_{00}-1$,
$\langle\Psi(t)|\sigma_X|\Psi(t)\rangle=2\textrm{Re}\rho_{10}$,
$\langle\Psi(t)|\sigma_Y|\Psi(t)\rangle=2\textrm{Im}\rho_{10}$. We
take the high frequency excitation in the form
$\widetilde{\varepsilon}(t)=\varepsilon_1\cos(\omega t)$ and
rewrite Hamiltonian (\ref{H_q}) in the eigenstate basis
$\Psi_\pm$:
\begin{equation}\label{Ham}
H_q=-\hbar\left(\Omega+\frac{\varepsilon_0\varepsilon_1}{\Omega}\cos(\omega
t)\right)\sigma_Z+\hbar\frac{\varepsilon_1\Delta}{\Omega}\cos(\omega
t)\sigma_X.
\end{equation}

From density matrix equation $ \label{rhot}
i\hbar\dot{\rho}(t)=\left[H_q,\rho(t)\right]$ we get  the
following set of equations for the matrix elements of $\rho$:

\begin{equation}\label{A1}
\frac{dA}{dt}=2\frac{\Delta\varepsilon_1}{\Omega}\cos(\omega t)B,
\end{equation}

\begin{equation}\label{B1}
\frac{dB}{dt}=-\frac{\Delta\varepsilon_1}{\Omega}\cos(\omega
t)(2A-1)-2\left(\Omega+\frac{\varepsilon_0\varepsilon_1}{\Omega}\cos(\omega
t)\right)C,
\end{equation}

\begin{equation}\label{C1}
\frac{dC}{dt}=2\left(\Omega+\frac{\varepsilon_0\varepsilon_1}{\Omega}\cos(\omega
t)\right)B,
\end{equation}
where $A=\rho_{00}$, $B=\textrm{Im}\rho_{10}$,
$C=\textrm{Re}\rho_{10}$.

These equations cannot be solved analytically, however, we can try
to predict the evolution of the quantities \textit{A, B, C} since
the Eqs. ~(\ref{A1},\ref{B1},\ref{C1}) are similar to those of a
free spin in external magnetic field in NMR, where  \textit{A} is
analogue of longitudinal magnetization $M_Z$, and \textit{B, C}
are analogues of transverse magnetizations $M_Y$ and $M_X$,
respectively. The only difference is that here the only detectable
quantity is $\sigma_Z$ component which is proportional to the
circulating current. Therefore, we could expect the evolution of
quantities \textit{A, B, C} under the influence of external high
frequency excitation is similar to that  in NMR. If the frequency
of external excitation $\omega$ is close to the gap frequency
$2\Omega$, the main harmonic of \textit{A} will be the Rabi
frequency
\begin{equation}\label{rabi1}
\omega_R=\sqrt{(\omega-2\Omega)^2+\Omega_R^2},
\end{equation}
where $\Omega_R$ depends on qubit parameters and on the amplitude
of excitation $\varepsilon_1$, while  the quantities \textit{B, C}
will oscillate with the gap frequency $2\Omega$, modulated by the
Rabi frequency: $B, C\approx\sin\Omega_Rt\sin2\Omega t$. If
$\varepsilon_1\ll\Omega$ we can estimate $\Omega_R$ from
Eqs.~(\ref{A1},\ref{B1},\ref{C1}) in rotating wave approximation.
We obtain
\begin{equation}\label{rabi2}
\Omega_R=\Delta\varepsilon_1/\Omega.
\end{equation}
 Since we want here to
treat the problem exactly we solved Eqs.
~(\ref{A1},\ref{B1},\ref{C1}) numerically.

For all computer simulations we take the following parameters of
qubit: $I_C=400$ nA, $L=15$ pH, $\alpha=0.8$, $g=100$,
$\Omega/2\pi=250$ MHz, $\varepsilon_0=\Delta$. The external high
frequency is equal to the gap: $\omega=2\Omega$. The excitation
amplitude $\varepsilon_1=0.28\Omega$ which ensures the Rabi
frequency $\Omega_R/2\pi=50$ MHz. In addition, we assume the qubit
is in its ground state in initial time $t_i=0$, so that
\textit{A}(0)=1, \textit{B}(0)=\textit{C}(0)=0. In all simulations
the time span is 10 $\mu$s from which the particular time windows
had been choose for the figures.

\begin{figure}[tbp]
\includegraphics[height=7cm,angle=-90]{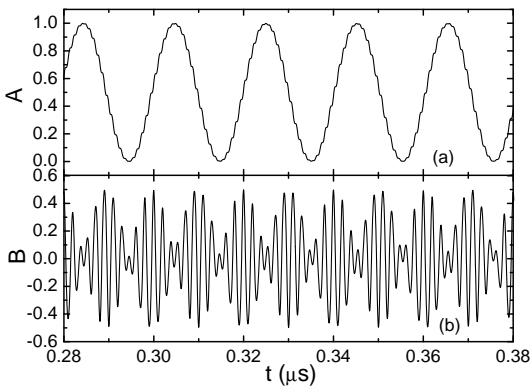}
\caption{Time evolution of \textit{A} and \textit{B} for qubit
without dissipation.} \label{fig2}
\end{figure}

The time evolution of \textit{A}(t) and \textit{B}(t) is shown on
Fig.~\ref{fig2}.

It is clearly seen that \textit{B} oscillates with gap frequency,
while the frequency of \textit{A} is almost ten times smaller:
(oscillation period of \textit{B}: $T_B\approx2\times10^{-9}$ s,
while the same quantity for \textit{A} is
$T_A\approx2\times10^{-8}$ s. The small distortions on \textit{A}
curve  are due to a strong deviation of excitation signal from
transverse rotating wave form, while \textit{B} curve is clearly
modulated with Rabi frequency $\Omega_R$ (Fig.~\ref{fig2}).

The crucial requirement for the proper operation of a qubit is the
preservation of phase coherence under influence of dissipative
environment. We include the environment effects phenomenologically
in Eqs. ~(\ref{A1},\ref{B1},\ref{C1}):

\begin{equation}\label{A2}
\frac{dA}{dt}=2\frac{\Delta\varepsilon_1}{\Omega}\cos(\omega
t)B-\frac{1}{T_r}(A-A_0),
\end{equation}

\begin{equation}\label{B2}
\frac{dB}{dt}=-\frac{\Delta\varepsilon_1}{\Omega}\cos(\omega
t)(2A-1)-2\left(\Omega+\frac{\varepsilon_0\varepsilon_1}{\Omega}\cos(\omega
t)\right)C-\frac{B}{T_d},
\end{equation}

\begin{equation}\label{C2}
\frac{dC}{dt}=2\left(\Omega+\frac{\varepsilon_0\varepsilon_1}{\Omega}\cos(\omega
t)\right)B-\frac{C}{T_d},
\end{equation}
where $\textit{A}_0$ is the equilibrium value of density matrix
$\textit{A}_0\equiv\rho_{00}^{eq}=\frac{1}{2}\left(1+\textrm{tanh}\frac{\hbar\Omega}{2k_BT}\right)$;
$T_r$ and $T_d$ are relaxation and dephasing times, respectively.
Here we have to consider density matrix elements as the quantities
averaged over environments degrees of freedom:
$\textit{A}=\langle|C_-(t)|^2|\rangle$,
$\textit{B}=\textrm{Im}\langle C_+(t)C^\ast_-(t)\rangle$,
$\textit{C}=\textrm{Re}\langle C_+(t)C^\ast_-(t)\rangle$. The Eqs.
~(\ref{A2},\ref{B2},\ref{C2}) are similar to well known
Bloch-Redfield equations in NMR\cite{Abragam}.

 For spin boson
model of coupling of a qubit to thermal bath these times have been
calculated for weak damping Ohmic spectrum by several authors
(see\cite{Leggett,Weiss} and references therein). Here for
estimations we take the expressions for $T_r$ and $T_d$
from\cite{Grifoni}
\begin{equation}\label{Tr}
\frac{1}{T_r}=\frac{1}{2}\left(\frac{\Delta}{\Omega}%
\right)^2J(\Omega) \coth\left(\frac{\hbar\Omega}{2k_BT}\right),
\end{equation}
\begin{equation}\label{Td}
\frac{1}{T_d}=\frac{1}{2T_r}+2\pi\eta\left(%
\frac{\varepsilon_0}{\Omega}\right)^2 \frac{k_BT}{\hbar},
\end{equation}
where dimensionless parameter $\eta$ reflects the strength of
Ohmic dissipation. $J(\Omega)$ is the spectral density of the bath
fluctuations at the gap frequency.

\begin{figure}[tbp]
\includegraphics[height=7cm,angle=-90]{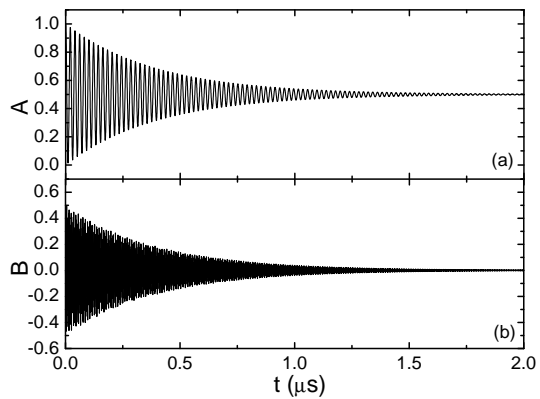}
\caption{Time evolution of \textit{A} and \textit{B} computed from
Eqs. (14)-(16). $T_r=50$ ms, $T_d=200$ ns.} \label{fig3}
\end{figure}

The decoherence is caused primarily by coupling of a solid state
based phase qubit to microscopic degrees of freedom in the solid.
Fortunately this intrinsic decoherence has been found to be quite
weak \cite{LinTian}. However, the external sources of decoherence
are more serious. Here we assume that the main source of
decoherence is the external flux noise. For our computer
simulations we take $T_r=50$ ms, $T_d=200$ ns\cite{Greenberg1}.
The evolution of \textit{A, B} found from Eqs.
~(\ref{A2},\ref{B2},\ref{C2}) is shown in Fig.~\ref{fig3}.

As is seen from the Fig.~\ref{fig3}, \textit{A} decays to 0.5
oscillating with Rabi frequency, while \textit{B (C)} decays to
zero. (Note: to be rigorous, the stable state solution for
\textit{A} is \cite{Abragam}
\begin{equation}\label{A0}
A\approx\frac{1}{2}+\frac{\rho_{00}^{eq}}{2(1+\Omega_R^2T_rT_d)}.
\end{equation}

However, as distinct from conventional NMR where
$\Omega_R^2T_rT_d\approx 1$, for our values of $\Omega_R$, $T_r$
and $T_d$ we get $\Omega_R^2T_rT_d\approx10^7$. For the same
reason the stable state oscillations of \textit{B} and \textit{C}
are quite small). It means the qubit density matrix becomes the
statistical mixture: $\rho_{00}\rightarrow 1/2$,
$\rho_{11}\rightarrow 1/2$, $\rho_{10}=\rho_{01}\rightarrow 0$ at
$t\rightarrow\infty$. Therefore, the noise from environment leads
to a delocalization: the system which initially is localized in
any state would be always delocalized at $t\rightarrow\infty$.

 This property has been first pointed
out in connection with noninvasive measurements of coherent
dynamics in quantum-dot systems\cite{Gurvitz} where one can find
an interesting discussion of how this behavior is related to the
quantum Zeno effect, and recently has been confirmed for a
quantum-dot qubit interacting with an environment and continuously
monitored by a tunnel-junction detector\cite{Gurvitz03}.

As is seen from  Eq. ~\ref{avcurr}), the delocalization leads to
the vanishing of the current in the qubit loop.

\section{The measuring of flux qubit with a tank circuit}
\label{bit-tank} Below we consider a continuous  measurements of a
qubit with a classical tank circuit which is weakly coupled to the
qubit via mutual inductance $M$. First we study the ideal case
when qubit and tank circuit are completely decoupled from their
environments, so that we may describe qubit + tank system by
Hamiltonian $H=H_q+H_T+H_{int}$, where

\begin{equation} \label{H_T}
H_T=\frac{Q^2}{2C_T}+\frac{\Phi^2}{2L_T}.
\end{equation}

In (\ref{H_T}) $C_\mathrm{T}$, and  $L_\mathrm{T}$ are capacitor
and inductor of a tank circuit; $Q$ is the charge at the
capacitor, $\Phi$ is the magnetic flux trapped by the inductor.
The tank-qubit interaction is described by Hamiltonian

\begin{equation}\label{H_int}
H_{int}=\lambda\widehat{I}_q\Phi,
\end{equation}
where qubit current operator $\widehat{I}_q$ is given in
(\ref{curr}), $\lambda=M/L_T$.

The equations for qubit + tank system are
\begin{equation}\label{A3}
\frac{dA}{dt}=2\frac{\Delta\varepsilon_1}{\Omega}\cos(\omega
t)B-2\lambda\eta I_C\frac{\Delta}{\hbar\Omega}B\Phi,
\end{equation}

\begin{eqnarray}\label{B3}
\frac{dB}{dt}=-\frac{\Delta\varepsilon_1}{\Omega}\cos(\omega
t)(2A-1)- \nonumber \\
2\left(\Omega+\frac{\varepsilon_0\varepsilon_1}{\Omega}\cos(\omega
t)\right)C+ \nonumber \\
2\lambda
I_C\frac{\eta}{\hbar\Omega}(2\varepsilon_0C+\Delta(2A-1))\Phi,
\end{eqnarray}

\begin{equation}\label{C3}
\frac{dC}{dt}=2\left(\Omega+\frac{\varepsilon_0\varepsilon_1}{\Omega}\cos(\omega
t)\right)B-2\lambda\eta I_C\frac{\varepsilon_0}{\hbar\Omega}B\Phi,
\end{equation}

\begin{equation}\label{flux_tank}
\frac{d\Phi}{dt}=\frac{Q}{C_T},
\end{equation}

\begin{equation}\label{Q}
\frac{dQ}{dt}=-\frac{\Phi}{L_T}-\lambda\eta I_CF[\xi(t)],
\end{equation}
where $\eta=\frac{\lambda(\alpha,g)}{2\pi}$. The function
$F[\xi(t)]$ in ~(\ref{Q}) stands for stochastic nature of the
measuring process. In accordance with von Neumann postulate the
outcome of a single measurement cannot be predicted
deterministically. When qubit is in a superposition of two
stationary states its wave function $\Psi(t)$ can be expressed in
the flux basis as $\Psi(t)=U\Psi_L+W\Psi_R$ where
\begin{equation}\label{U}
|U|^2=a_+^2+A(a_-^2-a_+^2)+2Ca_+a_-,
\end{equation}
\begin{equation}\label{W}
   |W|^2=b_+^2+A(b_-^2-b_+^2)+2Cb_+b_-
\end{equation}
with $|U|^2+|W|^2=1$.

The states $\Psi_L$, $\Psi_R$ have equal currents circulating in
opposite directions so that the outcome of the measurement (the
direction of a current circulation) can be predicted only
statistically with probability $|U|^2$ or $|W|^2$, respectively.
Following this reasoning we take $F[\xi(t)]$ in the form:

\begin{equation}\label{F}
F[\xi(t)]=\frac{|U|^2-\xi(t)}{||U|^2-\xi(t)|},
\end{equation}

where $\xi(t)$ generates random numbers from interval $[0,1]$.
Since $|U|^2\leq1$  the function $F[\xi(t)]$ takes two values:
$+1$, $-1$. It accounts for unpredictable nature of a single
measurement. If, for example, in some moment $t_i$
$|U(t_i)|^2>0.5$ it is more probable to find the clockwise than
counterclockwise direction of circulating current at this moment
of time in a single measurement. The actual value of the voltage
across the tank at some time $t_i$ is obtained as the average of
individual measurements over N different realizations of $\xi(t)$:

\begin{equation}\label{Real}
    V(t_i)=\frac{1}{NC_T}\sum_{j=1}^{j=N}Q(t_i,\xi_j).
\end{equation}

Therefore, our model accounts for stochastic back action influence
of the measuring device (tank circuit) on the qubit behavior. In
some sense the model resembles the probabilistic measurement
scheme described in\cite{Korotkov2, Korotkov} for detection of the
electron position in double quantum dot by measuring the current
through tunnel junction coupled to quantum double dot qubit.

\begin{figure}
\includegraphics[height=7cm, angle=-90]{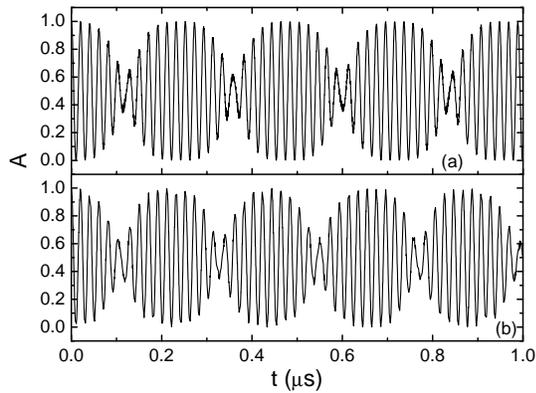}
\caption{Phase loss-free qubit coupled to a loss-free tank
circuit. Oscillations of \textit{A}. Deterministic case (a)
together with one realization (b) are shown. Small scale time
oscillations correspond to Rabi frequency.} \label{fig4a}
\end{figure}

\begin{figure}[tbp]
\includegraphics[height=7cm, angle=-90]{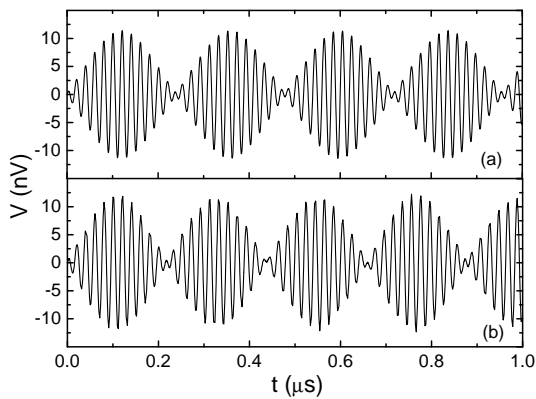}
\caption{Phase loss-free qubit coupled to a loss-free tank
circuit. Voltage across the tank. Deterministic case (a) together
with one realization (b) are shown.} \label{fig4b}
\end{figure}

Below, the tank circuit parameters are $L_T=50$ nH, $C_T=200$ pF,
so that the tank is tuned to 50 MHz. The inductive coupling to the
qubit $M=12.5$ pH that gives for the coupling parameter
$\lambda=2.5\times10^{-4}$. In addition, we take
$\varepsilon_0=\Delta$ so that
\begin{equation}\label{U1}
    |U|^2=\frac{1}{\sqrt{2}}\left(\frac{\sqrt{2}+1}{2}-A+C\right).
\end{equation}
 The results of computer simulations
of the equation set (\ref{A3},\ref{B3},\ref{C3},\ref{flux_tank},
\ref{Q}) are shown in Figs~\ref{fig4a}~and~\ref{fig4b}.

At every graph of the figures the results for one realization of
random number generator $\xi(t)$ are compared with the case when
we replaced $F[\xi(t)]$ in (\ref{Q}) with deterministic term
$(2\textit{A}-1-2C)/\sqrt{2}$, which means that the tank measures
the average current (\ref{avcurr}) in a qubit loop. As is seen
from the Fig.~\ref{fig4a}, \textit{A} oscillates with Rabi
frequency. The voltage across tank circuit oscillates also with
Rabi frequency which is equal to  50 MHz in our case
(Fig.~\ref{fig4b}) which is modulated with the lower frequency the
value of which is about 5 MHz.

It is worth to note the interesting feature of the result: though
the system is free from dissipation the voltage across the tank is
not saturated (the voltage amplitude is about 10 nV,
Fig.~\ref{fig4b}). Although \textit{A} oscillates at resonance
frequency of the tank, the saturation is not reached. A simple
estimations show that at the tank resonance the saturated value of
the voltage is about 75 nV.
 We have found the effect is due to
the large value of a coupling constant $\lambda$. The simulations
show the full saturation is reached with $\lambda\approx10^{-8}$,
however, then the voltage is quite small to be detected. These
results are valid exactly for loss-free tank circuit, however we
may within simulation time span (10 $\mu$s) extrapolate them for
tank with high quality factor, say $Q_T>1000$.

\begin{figure}[tbp]
\includegraphics[height=7cm, angle=-90]{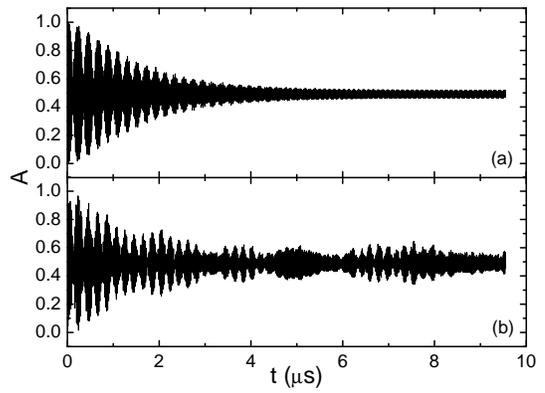}
\caption{Phase loss-free qubit coupled to a dissipative tank
circuit. The evolution of \textit{A} exhibits modulation of  Rabi
oscillations with lower frequency. Deterministic case (a) together
with one realization (b) are shown.} \label{fig5}
\end{figure}

\begin{figure}[tbp]
\includegraphics[height=7cm, angle=-90]{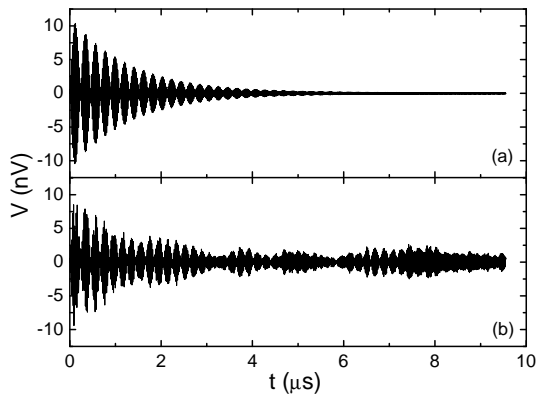}
\caption{Phase loss-free qubit coupled to a dissipative tank
circuit. The voltage across the tank exhibits  modulation of Rabi
frequency. Deterministic case (a) together with one realization
(b) are shown.} \label{fig6}
\end{figure}

Now we want to account for the damping in the tank circuit. We
replace Eq. \ref{Q} with
\begin{equation}\label{Q1}
\frac{dQ}{dt}=-\frac{\Phi}{L_T}-\frac{\omega_0}{Q_T}Q-\lambda\eta
I_CF[\xi(t)].
\end{equation}
For the simulations we take tank quality $Q_T=100$. The results of
simulations of equation set (\ref{A3},\ref{B3}, \ref{C3},
\ref{flux_tank}, \ref{Q1}) are shown for \textit{A} in
Figs.~\ref{fig5}, and for the voltage in Figs.~\ref{fig6}.

It is worthwhile to note that though the qubit is uncoupled from
its own environment, nevertheless, the current in a qubit loop and
the voltage across the tank decay. The quantities \textit{B} and
\textit{C} which are not shown here oscillate without damping with
the frequency of excitation with the amplitude equal to 0.5 .

 Finally, we consider the case when the qubit and the tank are
 coupled to their own environments. The corresponding set of
 equations are Eqs. ~(\ref{flux_tank},\ref{Q1}) and following equations:

\begin{equation}\label{A4}
\frac{dA}{dt}=2\frac{\Delta\varepsilon_1}{\Omega}\cos(\omega
t)B-2\lambda\eta
I_C\frac{\Delta}{\hbar\Omega}B\Phi-\frac{1}{T_r}(A-A_0),
\end{equation}

\begin{eqnarray}\label{B4}
\frac{dB}{dt}=-\frac{\Delta\varepsilon_1}{\Omega}\cos(\omega
t)(2A-1)-\nonumber\\
2\left(\Omega+\frac{\varepsilon_0\varepsilon_1}{\Omega}\cos(\omega
t)\right)C+\nonumber\\
2\lambda
I_C\frac{\eta}{\hbar\Omega}(2\varepsilon_0C+\Delta(2A-1))\Phi-\frac{B}{T_d},
\end{eqnarray}

\begin{equation}\label{C4}
\frac{dC}{dt}=2\left(\Omega+\frac{\varepsilon_0\varepsilon_1}{\Omega}\cos(\omega
t)\right)B-2\lambda\eta
I_C\frac{\varepsilon_0}{\hbar\Omega}B\Phi-\frac{C}{T_d}.
\end{equation}

\begin{figure}[tbp]
\includegraphics[height=7cm, angle=-90]{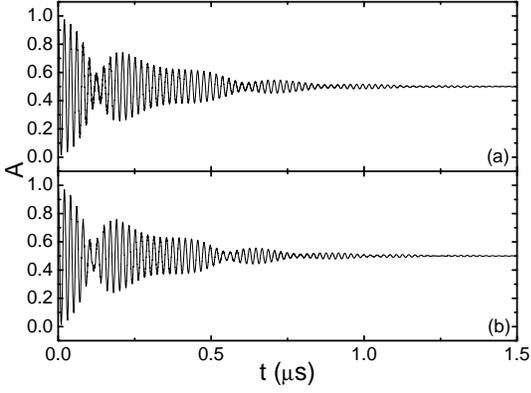}
\caption{Phase dissipative qubit coupled to a loss-free tank
circuit $(1/Q_T=0)$. The evolution of \textit{A}. Deterministic
case (a) together with one realization (b) are shown.}
\label{fig7a}
\end{figure}

\begin{figure}[tbp]
\includegraphics[height=7cm, angle=-90]{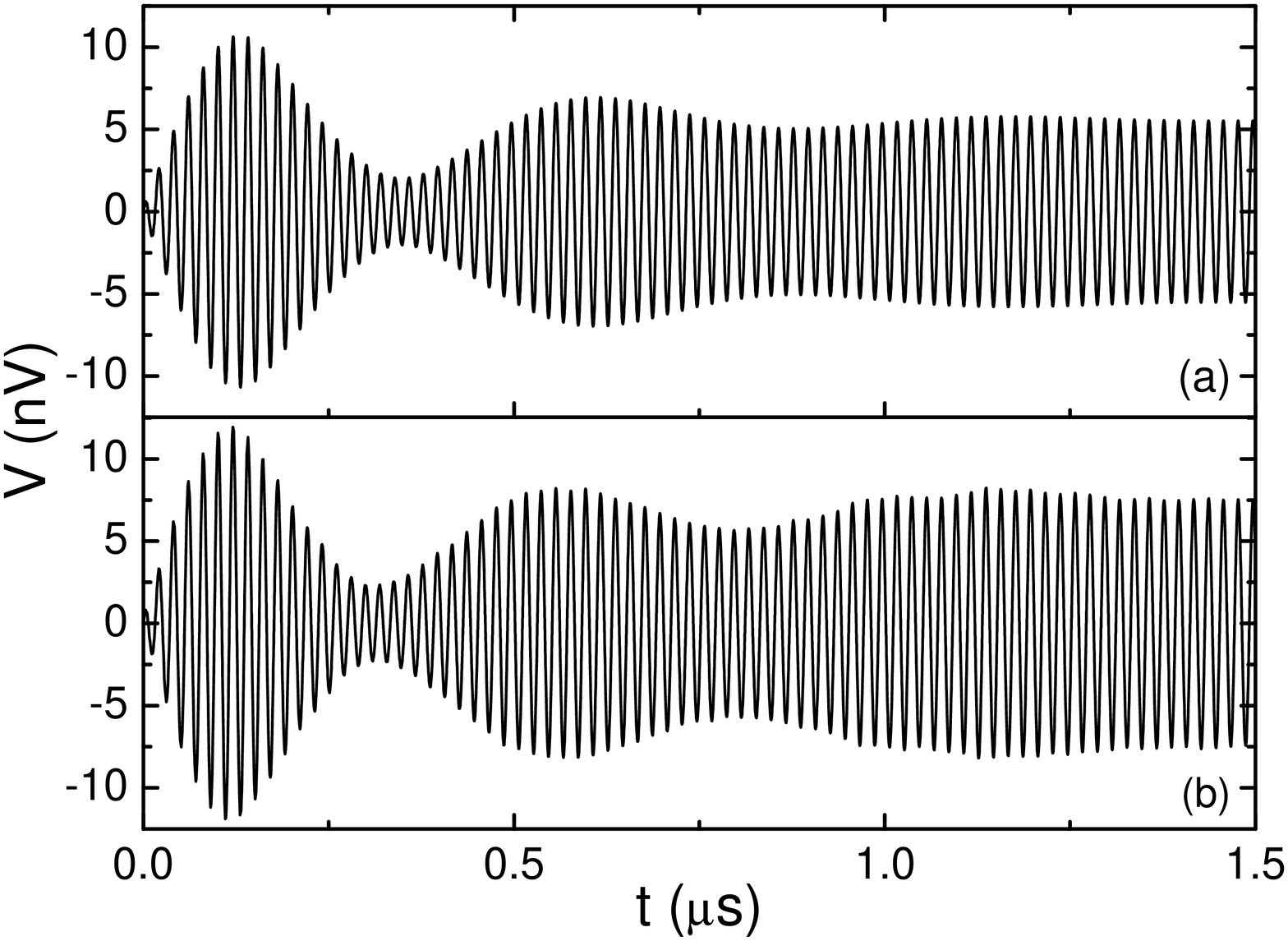}
\caption{Phase dissipative qubit coupled to a loss-free tank
circuit $(1/Q_T=0)$. The voltage across tank circuit.
Deterministic case (a) together with one realization (b) are
shown.} \label{fig7b}
\end{figure}

\begin{figure}[tbp]
\includegraphics[height=7cm, angle=-90]{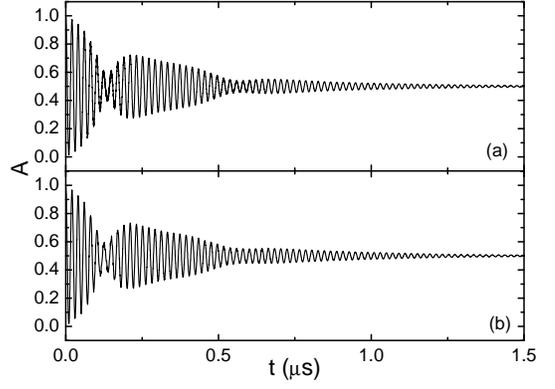}
\caption{Phase dissipative qubit coupled to a dissipative tank
circuit $(Q_T=100)$. The evolution of \textit{A}. One realization
(lower graph) together with deterministic case (upper graph) are
shown.} \label{fig8a}
\end{figure}

\begin{figure}[tbp]
\includegraphics[height=7cm, angle=-90]{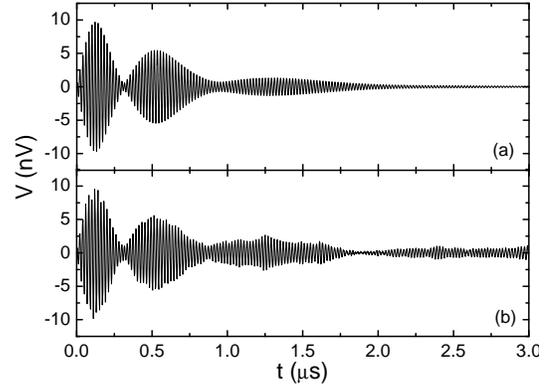}
\caption{Phase dissipative qubit coupled to a dissipative tank
circuit $(Q_T=100)$. The voltage across tank circuit. One
realization (lower graph) together with deterministic case (upper
graph) are shown.} \label{fig8b}
\end{figure}

\begin{figure}[tbp]
\includegraphics[height=7cm, angle=-90]{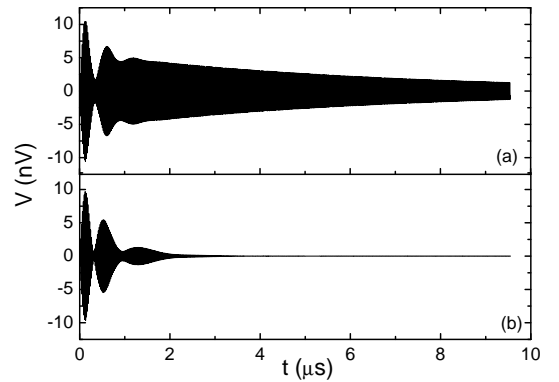}
\caption{Phase dissipative qubit coupled to a dissipative tank
circuit. The voltage across the tank for deterministic case.
Quality factors: $Q_T=1000$ (upper graph) and $Q_T=100$ (lower
graph).}
 \label{fig9}
\end{figure}

The results of simulations of the equation set
(\ref{A4},\ref{B4},\ref{C4},\ref{flux_tank},\ref{Q1}) are shown in
Figs.~\ref{fig7a},~\ref{fig7b},~\ref{fig8a},~\ref{fig8b}, and
Fig.~\ref{fig9}.

\begin{figure}[tbp]
\includegraphics[height=7cm, angle=-90]{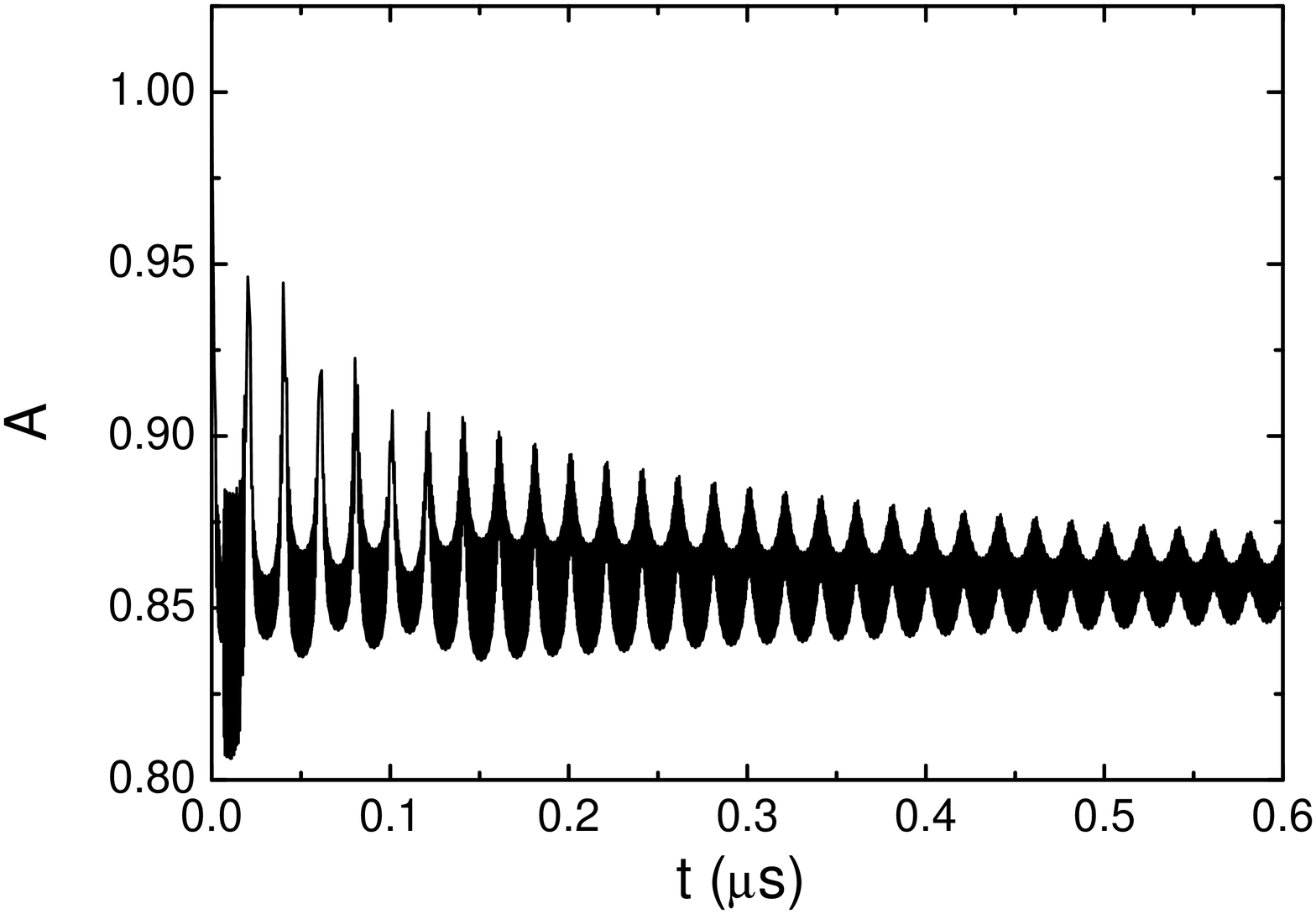}
\caption{Loss-free qubit coupled to a dissipative tank circuit
$(Q_T=100)$, $\lambda=2.5\times10^{-2}$. The evolution of
\textit{A}.  A distance between neighbor jumps is equal to Rabi
period. Deterministic case is shown.} \label{fig10a}
\end{figure}

\begin{figure}[tbp]
\includegraphics[height=7cm, angle=-90]{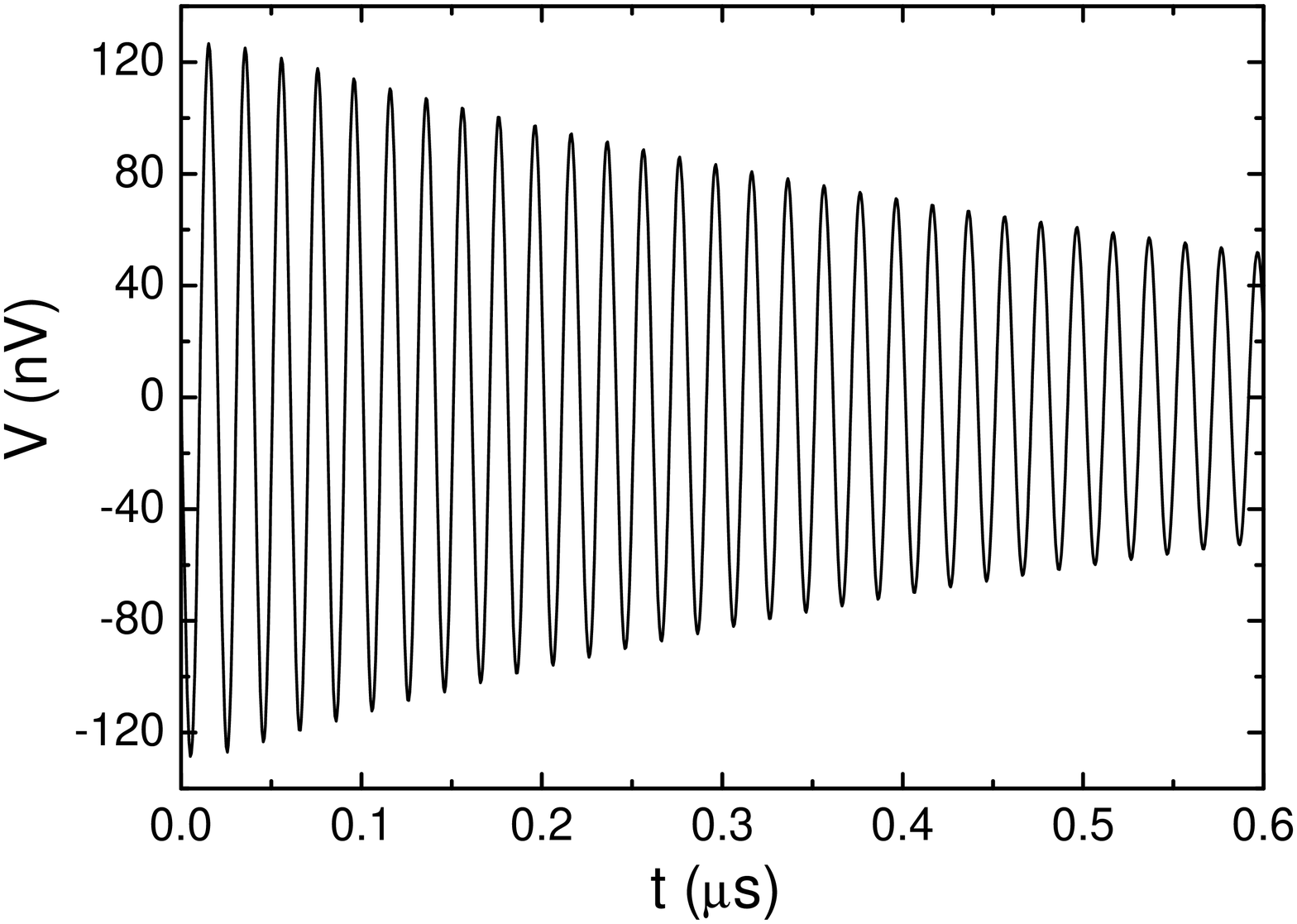}
\caption{Loss-free qubit coupled to a dissipative tank circuit
$(Q_T=100)$, $\lambda=2.5\times10^{-2}$.  The evolution of the
voltage across the tank for the deterministic case.}
\label{fig10d}
\end{figure}

 From Fig.~\ref{fig7a} and Fig.~\ref{fig8a} we see that \textit{A}
is almost unaffected by the value of $Q_T$. Its decay is defined
primarily by the shortest time $T_d$. The decay time of the
voltage is defined by the value of $Q_T$, since tank circuit decay
time $2Q_T/\omega_T$ is much longer than the dephasing time $T_d$
(Figs.~\ref{fig7b},~\ref{fig8b},~\ref{fig9}). It is advantageous
from the point of experiment, since we can measure Rabi
oscillations much longer than the dephasing time $T_d$. However,
from the other hand, the maximum value of the voltage amplitude is
almost independent of $Q_T$ (see Fig. ~\ref{fig9}). We have shown
in\cite{Greenberg1} that for carefully made electronics the
voltage noise at the input of preamplifier could be on the order
of 10 nV-20 nV. As is seen from Fig.~\ref{fig8b} and especially,
from Fig.~\ref{fig9}, during the first microsecond the
signal-to-noise ratio is about 0.5. The signal can be recovered
with the well known in NMR pulsed technique with subsequent
Fourier processing. However, since here the pulse filling
frequency is uncoupled from the frequency of the signal to be
detected, it is not necessary to keep the pulse width shorter than
the decaying time of the signal. As our results show, the pulse
duration of about 2 $\mu$s is adequate for the measurements.

In conclusion we want to show the effect of qubit evolution as the
coupling between the qubit and the tank is increased. We
numerically solved the system consisting of the loss-free qubit
coupled to the dissipative tank circuit. The system is described
by Eqs.~(\ref{A2},\ref{B2},\ref{C2},\ref{flux_tank}) and Eq.
~(\ref{Q1}). For the simulations we take the coupling parameter
$\lambda=2.5\times10^{-2}$. The results of simulations are shown
on Figs.~\ref{fig10a},~\ref{fig10d} for deterministic case. As is
seen from the Figs.~\ref{fig10a} during Rabi period the quantity
\textit{A} became partially frozen at some level. At the endpoints
of this period the system tries to escape to another level of
\textit{A}. Between the endpoints of Rabi period \textit{A}
oscillates with a high frequency which is about 10 GHz in our
case. As expected, the evolution of \textit{B} is suppressed
approximately by a factor of ten below its free evolution
amplitude which is equal to 0.5.  As we show below, the strong
coupling completely destroys the phase coherence between qubit
states, nevertheless the voltage across the tank oscillates with
Rabi frequency. Its amplitude is considerably increased and it
does not reveal any peculiarities associated with the frozen
behavior of \textit{A} (Fig.~\ref{fig10d}).

\section{Qubit wave function}
In conclusion we want to study the effect of a coupling between
qubit and the tank on the qubit state, in particular, on phase
coherence between basis states of the qubit. It is necessary to
note that our \textit{measurement} is not the \textit{measurement}
in the sense of Neumann. We are interested only in the voltage
amplitude in the tank but not in the state of the qubit: we did
not solve Schrodinger equations for $C_-=|C_-|\exp (i\varphi_-)$,
$C_+=|C_+|\exp (i\varphi_+)$ but for their products $A=|C_-|^2$,
$B=|C_-||C_+|\sin(\varphi_+-\varphi_-)$,
$C=|C_-||C_+|\cos(\varphi_+-\varphi_-)$. Nevertheless, we can
check to what extent the qubit can be described by the wave
function in case of its interaction with a tank. Evidently, free
qubit must have definite wave function at any instant of time. It
means the conservation of phase coherence the condition for which
can be expressed in terms of our quantities as:
\begin{equation}\label{coh}
    \frac{B^2+C^2}{|C_-|^2|C_+|^2}=1.
\end{equation}

\begin{figure}[tbp]
\includegraphics[height=7cm, angle=-90]{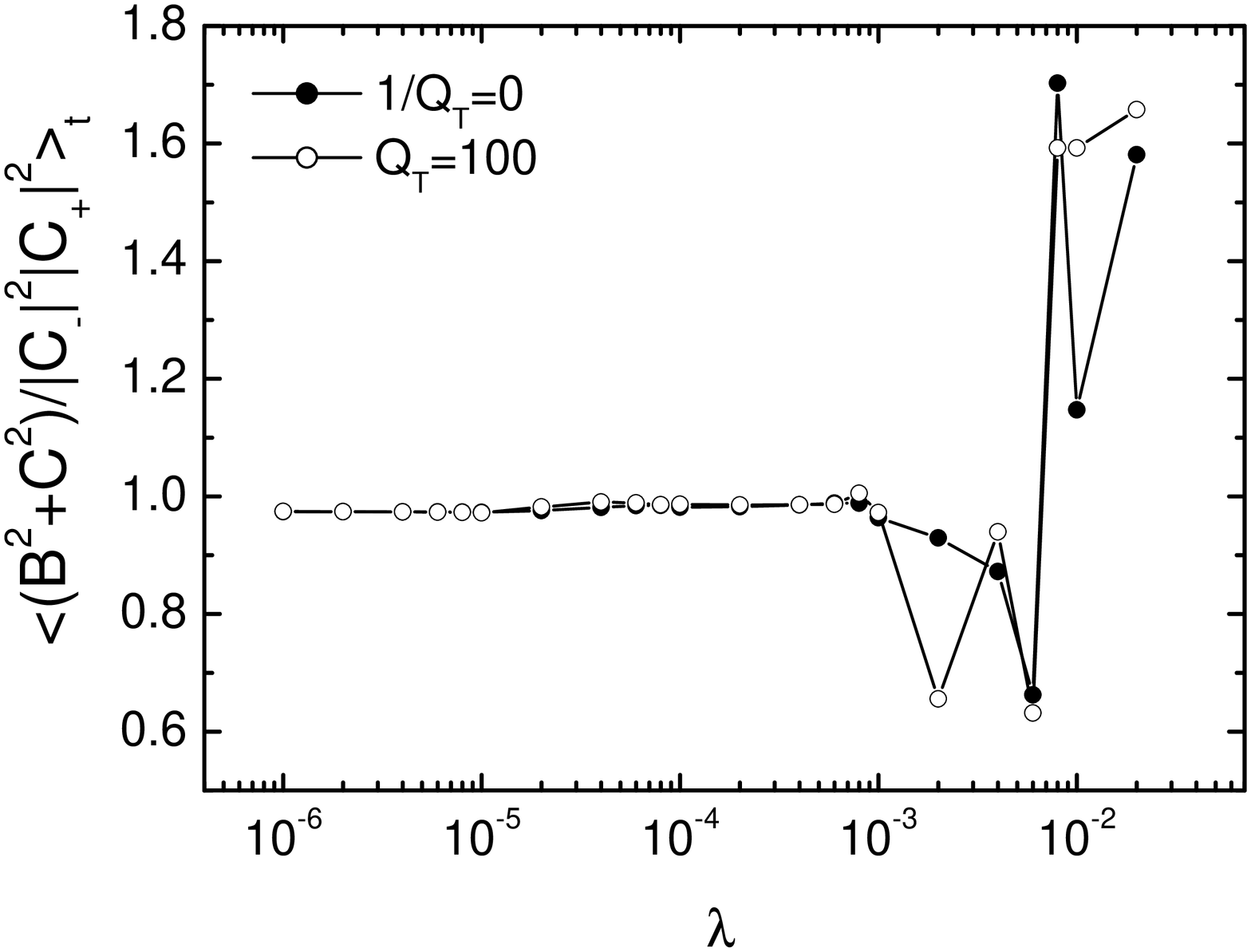}
\caption{The condition of phase coherence (\ref{coh}) vs coupling
strength $\lambda$.} \label{figcoh}
\end{figure}

If we switch on the interaction with a tank we may not, strictly
speaking, consider qubit as having definite wave function.
However, if the interaction is rather weak the qubit wave function
could be well defined. We showed before that for relatively weak
coupling the dissipation resulted in quenching \textit{A} to the
0.5 level (see Figs.~\ref{fig5}, \ref{fig7a}, \ref{fig8a}). That
means $|C_+|=|C_-|\rightarrow\frac{1}{\sqrt{2}}$. However, as is
seen from Fig.~\ref{figcoh}, the condition of phase coherence is
still valid up to $\lambda\approx10^{-3}$.

As the coupling is increased further the qubit wave function is
completely destroyed. The quantity \textit{A} is quenched to
approximately 0.85 (Fig.~\ref{fig10a}). That is
$|C_-|\approx0.92$. It might seem that we have here so called Zeno
effect- as if qubit state is frozen in its ground state. However,
in case of a strong coupling it is not correct to say about wave
function of the qubit alone. This is shown in Fig.~\ref{figcoh}
where for $\lambda>10^{-2}$ the phase coherence is seen to be
completely lost .

\section{Conclusion}
In all computer simulations we systematically compared different
realizations with the case when we replaced $F[\xi(t)]$ in
(\ref{Q}) with deterministic term $(2\textit{A}-1-2C)/\sqrt{2}$.
We have found that within a decaying time all realizations and
deterministic case give almost identical results. That is why in
all corresponding graphs we compared deterministic case with the
only realization. A clear difference appears only at the tails
where \textit{A} is close to 0.5 and  \textit{C}is rather small.
This is because at the tails the random number generator every
moment of time changes the sign of the current with a high
probability, while within a decaying time where \textit{A}
undergoes oscillations the sign of the current for one half of
period of oscillations of \textit{A} is conserved with a high
probability.

Throughout the paper we stress the similarity between the
qubit+tank system and NMR, however, we have to be aware of the
main difference. In NMR the back action of the tank circuit on the
sample under study is neglected. It is justified by the fact that
the tank is coupled to macroscopic number of two-level systems
(1/2-spin particles). The coupling to the individual particle is
rather small, so that a reasonable signal level at the tank is
obtained at the expense of enormous number of the coupled
particles. However, when the tank is coupled to a single two-level
system the account for back action is necessary. It leads to the
main quantitative difference from NMR. In order to keep the noise
from the tank as small as possible, the quality factor $Q_T$
should be taken as high as possible from technological point of
view (in our simulations we take $Q_T=100$ only in order to save
the simulation time). However, the signal amplitude weakly depends
on $Q_T$ being at best at the level of noise. Nevertheless, it is
not difficult to recover the signal with the aid of the methods of
signal processing which are used in high resolution NMR.
Therefore, the results of our simulations clearly show that we can
detect Raby oscillations of the voltage across tank circuit
coupled to the qubit with the  pulsed Fourier technique which is
well known in NMR.

\section{Acknowledgements}
I should like to express my gratitude to E. Il'ichev and M.
Grajcar for many fruitful discussions on various experimental and
theoretical aspects concerning the problems considered in the
paper. I gratefully acknowledge A. Izmalkov for his help with
computer simulations. I also want to express my appreciation to A.
Maassen van den Brink for critical reading of manuscript and
useful comments. The work was supported by INTAS Program of EU
under grant 2001-0809.

\end{document}